\documentclass{elsart}
\usepackage{ifpdf}
\usepackage{graphicx,natbib,amssymb,lineno}
\ifpdf
\usepackage[%
  pdftitle={Instructions for use of the document class
    elsart},%
  pdfauthor={Simon Pepping},%
  pdfsubject={The preprint document class elsart},%
  pdfkeywords={instructions for use, elsart, document class},%
  pdfstartview=FitH,%
  bookmarks=true,%
  bookmarksopen=true,%
  breaklinks=true,%
  colorlinks=true,%
  linkcolor=blue,anchorcolor=blue,%
  citecolor=blue,filecolor=blue,%
  menucolor=blue,pagecolor=blue,%
  urlcolor=blue]{hyperref}
\else
\usepackage[%
  breaklinks=true,%
  colorlinks=true,%
  linkcolor=blue,anchorcolor=blue,%
  citecolor=blue,filecolor=blue,%
  menucolor=blue,pagecolor=blue,%
  urlcolor=blue]{hyperref}
\fi

\makeatletter
\def\elsartstyle{%
    \def\normalsize{\@setfontsize\normalsize\@xiipt{14.5}}
    \def\small{\@setfontsize\small\@xipt{13.6}}
    \let\footnotesize=\small
    \def\large{\@setfontsize\large\@xivpt{18}}
    \def\Large{\@setfontsize\Large\@xviipt{22}}
    \skip\@mpfootins = 18\p@ \@plus 2\p@
    \normalsize
} \@ifundefined{square}{}{\let\Box\square} \makeatother

\def\file#1{\texttt{#1}}

\begin{document}
\begin{frontmatter}

\title{Resonant energy conversion of 3-minute intensity oscillations into
Alfv{\'e}n waves in the solar atmosphere}

\author{D. Kuridze and T.V. Zaqarashvili}
\address{Georgian National Astrophysical Observatory (Abastumani Astrophysical Observatory), Al. Kazbegi ave.
2a, 0160 Tbilisi, Georgia}

\ead{dato.k@genao.org} \ead{temury@genao.org}

\begin{abstract}
Nonlinear coupling between 3-minute oscillations and Alfv{\'e}n
waves in the solar lower atmosphere is studied. 3-minute
oscillations are considered as acoustic waves trapped in a
chromospheric cavity and oscillating along transversally
inhomogeneous vertical magnetic field. It is shown that under the
action of the oscillations the temporal dynamics of Alfv{\'e}n waves
is governed by Mathieu equation. Consequently, the harmonics of
Alfv{\'e}n waves with twice period and wavelength of 3-minute
oscillations grow exponentially in time near the layer where the
sound and Alfv{\'e}n speeds equal i.e. $c_s \approx v_A$. Thus the
3-minute oscillations are resonantly absorbed by pure Alfv{\'e}n
waves near this resonant layer. The resonant Alfv{\'e}n waves may
penetrate into the solar corona taking energy from the chromosphere.
Therefore the layer $c_s \approx v_A$ may play a role of {\it energy
channel} for otherwise trapped acoustic oscillations.
\end{abstract}

\end{frontmatter}

\section{Introduction}
\label{intro}

Waves play an important role in the dynamics of the solar
atmosphere. They are believed as carriers of photospheric energy
into the corona leading to plasma heating. Waves are intensively
observed in the solar chromosphere and corona by SOHO (Solar and
Heliospheric Observatory) and TRACE (Transition Region and Coronal
Explorer) (Nakariakov \& Verwichte 2005). Dominant intensity
oscillations in a chromospheric network have a period of
$\sim$3-min. The 3-min oscillations are believed to be either
standing waves in a chromospheric cavity (Leibacher \& Stein 1981)
or propagating waves, which arise as the response of the atmosphere
to a general disturbance set at its base (Fleck \& Schmitz 1991).
Coupling of these oscillations to the Alfv{\'e}n waves is a crucial
process in both cases, as the Alfv{\'e}n waves may easily pass
through the transition region taking the energy into the corona.

On the other hand, recent numerical (Rosenthal et al., 2002),
analytical (Zaqarashvili \& Roberts, 2006) and observational
(Muglach et al., 2005) studies point out the importance of
$\beta\sim 1$ region of the solar atmosphere in the wave coupling
phenomena.

Here we consider the 3-min oscillations as the standing acoustic
waves oscillating along an uniform vertical magnetic field and study
their weakly nonlinear coupling to Alfv{\'e}n waves near the
chromospheric $\beta\sim 1$ regions.

\section[]{The wave coupling}

We use Cartesian coordinate system $(x,y,z)$, where spatially
inhomogeneous (along the $x$ axis) magnetic field is directed along
the $z$ axis, i.e. $B_0=(0,0,B_z(x))$. Unperturbed plasma pressure
$p_0(x)$ and density $\rho_0(x)$ are also assumed to vary along $x$.
In the equilibrium, magnetic and hydrodynamic pressures satisfy the
transverse balance, i.e. $p_0(x) + {{B^2_z(x)}/{8\pi}}=const$.
Plasma $\beta$ is defined as $\beta = {{8\pi p_0(x)}/{B^2_z(x)}}=
{{2c^2_s}/{\gamma v^2_A(x)}}$, where $c_s=\sqrt{\gamma p_0/\rho_0}$
and $v_A(x)=B_z/\sqrt{4 \pi \rho_0}$ are the sound and Alfv{\'e}n
speeds respectively and $\gamma$ is the ratio of specific heats. For
simplicity, temperature is suggested to be homogeneous so that the
sound speed does not depend on the $x$ coordinate.

We consider wave propagation along the $z$ axis (thus along the
magnetic field) and wave polarisation in $yz$ plane. Then in this
case only sound and linearly polarised Alfv{\'e}n waves arise. The
velocity component of sound wave is polarised along the $z$ axis and
the velocity component of Alfv{\'e}n wave is polarised along the y
axis. Then the ideal MHD equations can be written as
\begin{equation}
{{{\partial b_y}}\over {\partial t}} + u_z{{{\partial b_y}}\over
{\partial z}} = - b_y{{{\partial u_z}}\over {\partial z}} +
B_z(x){{{\partial u_y}}\over {\partial z}},
\end{equation}
\begin{equation}
{\rho}{{{\partial u_y}}\over {\partial t}} + {\rho}u_z{{{\partial
{u_y}}}\over {\partial z}} = {{B_z(x)}\over {4\pi}}{{\partial
b_y}\over {\partial z}},
\end{equation}
\begin{equation}
{{{\partial {\rho}}}\over {\partial t}}= - {\rho}{{{\partial
{u_z}}}\over {\partial z}} - u_z{{{\partial {\rho}}}\over {\partial
z}},
\end{equation}
\begin{equation}
{\rho}{{{\partial u_z}}\over {\partial t}} + {\rho}u_z{{{\partial
{u_z}}}\over {\partial z}} = - {{{\partial p}}\over {\partial z}} -
{{\partial}\over {\partial z}}{{b^2_y}\over {8\pi}},
\end{equation}
\begin{equation}
{{{\partial p}}\over {\partial t}} = - {\gamma}p{{{\partial
{u_z}}}\over {\partial z}} - u_z{{{\partial p}}\over {\partial z}},
\end{equation}
where $p=p_0 + p_1$ and $\rho={\rho}_0 + \rho_1$ denote the total
(unperturbed plus perturbed) hydrodynamic pressure and density,
$u_y$ and $u_z$ are velocity perturbations (of the Alfv{\'e}n and
sound waves, respectively), $b_y$ is the perturbation in the
magnetic field. Note, that the $x$ coordinate stands as a parameter
in these equations.

Eqs. (1)-(5) describe the fully nonlinear behavior of sound and
linearly polarized Alfv{\'e}n waves propagating along the magnetic
field. However the sound waves may be trapped in a chromospheric
cavity between the photosphere and transition region leading to
standing patterns. Therefore here we consider the sound waves
oscillating along the $z$ axis and bounded at $z=0$ and $z=l$
points. Thus there are the standing patterns
\begin{equation}
u_z=v(t)\sin(k_nz),\,\,\,\,\,\,  \rho_1 = {\tilde
\rho}(t)\cos(k_nz),
\end{equation} where $k_n$ is the wavenumber of sound wave such that
$ k_nl={{2\pi l}/{\lambda_n}}= n\pi$, so $ {l/{\lambda_n}}=n/2$,
where $n=1,2...$ denotes the order of corresponding harmonics.

It is natural that almost whole oscillation energy in bounded
systems is stored in a fundamental harmonic. Therefore here we
consider the first ($n=1$) harmonic of acoustic oscillations,
however the same can be applied to harmonics with arbitrary $n$.
Recently, Zaqarashvili and Roberts (2006) found that the harmonics
of acoustic and Alfv{\'e}n waves are coupled when the wave length of
acoustic wave is half of Alfv{\'e}n wave one. Therefore let express
the Alfv{\'e}n wave components as
\begin{equation}
u_y=u(t)\sin(k_Az),\,\,\,\,\,\,  b_y=b(t)\cos(k_Az),
\end{equation}
where $k_A$ is the wavenumber of the Alfv{\'e}n waves and the
condition $k_1=2k_A$ is satisfied.

Then the substitution of expressions (6)-(7) into Eqs. (1)-(5) and
averaging with $z$ over the distance $(0,l)$ leads to
\begin{equation}
{{{\partial b}}\over {\partial t}} = k_AB_0u + k_Avb,
\end{equation}
\begin{equation}
{{{\partial u}}\over {\partial t}} = - {{k_AB_0}\over {4\pi\rho_0}}b
- k_Auv,
\end{equation}
\begin{equation}
{{{\partial v}}\over {\partial t}} = {{k_1c^2_s}\over
{\rho_0}}{\tilde \rho} + {{k_A}\over {8\pi\rho_0}}b^2,
\end{equation}
\begin{equation}
{{{\partial {\tilde \rho}}}\over {\partial t}} = - \rho_0k_1v.
\end{equation}

Here Eqs. (10)-(11) describe the time evolution of acoustic
oscillation forced by ponderomotive force of Alfv{\'e}n waves, while
Eqs. (8)-(9) governs the dynamics of Alfv{\'e}n waves forced by
acoustic waves in a parametric way. It must be noted that the
coupling between sound and Alfv{\'e}n waves at $\beta \sim 1$ has
been recently studied by Zaqarashvili \& Roberts (2006). They
consider the general case of propagating waves and show an alternate
energy exchange between the waves during the propagation. Here we
consider the coupling between standing patterns of the waves, which
is the particular case of their results.

Substitution of $u$ from Eq. (8) into Eq. (9) and neglecting of all
third order terms leads to the second order differential equation
for Alfv{\'e}n waves

\begin{equation}
{{{\partial^2 b}}\over {\partial t^2}} + k^2_Av^2_A\left [1 -
{{2\tilde \rho}\over {\rho_0}} \right ] b= 0.
\end{equation}

This equation reflects the parametric influence of standing acoustic
wave due to the density variation. Then the particular time
dependence of density perturbation determines the type of equation
and consequently its solutions. If we consider the initial amplitude
of Alfv{\'e}n waves smaller than the amplitude of acoustic waves,
then the term with $b^2$ in Eq. (10) can be neglected. Physically it
means that the back reaction of Alfv{\'e}n waves due to the
ponderomotive force is small. Then the solution of Eqs. (10)-(11) is
just harmonic function of time $ {\tilde \rho} = \alpha \rho_0
\cos(\omega_1 t)$, where $\omega_1$ is the frequency of the first
harmonic of standing acoustic wave and $\alpha>0$ is the relative
amplitude. Here we consider the small amplitude acoustic waves
$\alpha \ll 1$, so the nonlinear steepening due to the generation of
higher harmonics is negligible. Then the substitution of this
expression into Eq. (12) leads to Mathieu equation
\begin{equation}
{{{\partial^2 b}}\over {\partial t^2}} + k^2_Av^2_A\left [1 -
{{2\alpha}}\cos(\omega_1 t) \right ] b= 0.
\end{equation}

The solution of this equation with the frequency of ${\omega_1}/2$
has an exponentially growing character, thus the main resonant
solution occurs when
\begin{equation}
{\omega_A}= v_Ak_A ={{\omega_1}\over 2},
\end{equation}
where $\omega_A$ is the frequency of Alfv{\'e}n waves. Since
$k_A=k_1/2$, the resonance takes place when $v_A =c_s$. Since the
Alfv{\'e}n speed $v_A(x)=B_0(x)/\sqrt{4 \pi \rho_0(x)}$ is a
function of the $x$ coordinate, then this relation will be satisfied
at particular locations along the $x$ axis. Therefore the acoustic
oscillations will be resonantly transformed into Alfv{\'e}n waves
near this layer. We call this region {\it swing layer} (see similar
consideration in Shergelashvili et al. 2005).

Under the resonant condition (14) the solution of Eq. (13) is
\begin{equation}
{b}(t)=b_0\exp{\left ({{\left |{\alpha}{\omega_1}\right |}\over 4}t
\right )}\left [{\cos}{{\omega_1}\over 2}t - {\sin}{{\omega_1}\over
2}t \right ],
\end{equation}
where $b_0= b(0)$.

The solution (15) has a resonant character within the frequency
interval ${\left |{\omega_A} - {{\omega_1}/ 2} \right |}<{\left
|{{\alpha \omega_1}/ 2} \right |}$. This expression can be rewritten
as ${\left |{{v_A}/ c_s} - 1 \right |}<{\left |{{\alpha }} \right
|}$. Thus the thickness of the resonant layer depends on the
acoustic wave amplitude. Therefore the acoustic oscillations are
converted into Alfv{\'e}n waves not only at the surface $v_A =c_s$
but in the region where $c_s \left (1 - {{\alpha }}\right )< v_A <
c_s \left (1 + {{\alpha }}\right )$. Thus the resonant layer can be
significantly wider for stronger amplitude acoustic oscillations.
Note, that the resonant Alfv{\'e}n waves expressed by Eqs. (7) and
(15) are standing patterns with the velocity node at the bottom
boundary ($z=0$) and antinode at the top boundary ($z=l$); the wave
length of  Alfv{\'e}n waves is twice than that of the acoustic
oscillations due to the condition $k_A=k_1/2$. Therefore the
oscillation of magnetic field lines at the upper boundary may excite
the waves in an external plasma, which carry energy away.

\begin{figure}
\begin{center}
   \includegraphics[width=10cm]{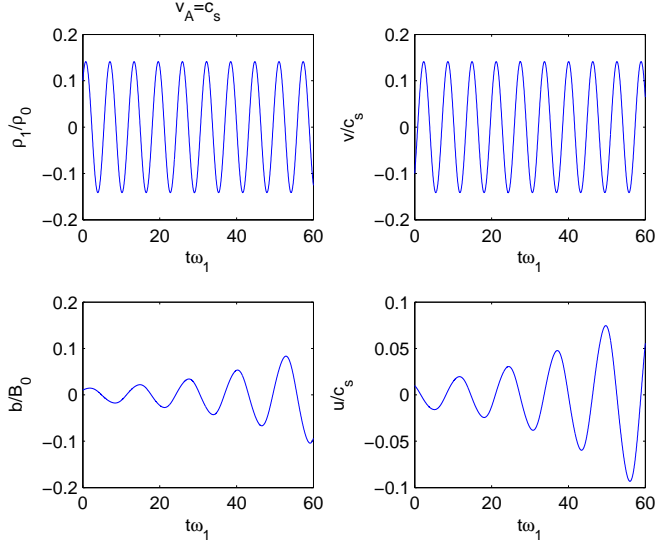}
      \caption{Numerical simulations of wave conversion in $c_s=v_A$ region. Upper panels
      show the density and velocity components of standing acoustic
      oscillations, while lower panels show the magnetic field and
      velocity component of Alfv{\'e}n waves. Relative amplitude of
      acoustic oscillations is 0.1. Alfv{\'e}n waves with twice the
      period of acoustic oscillations show rapid exponential increase in time.}
        \end{center}
         \label{FigVibStab}
   \end{figure}

It must be noted that the Alfv{\'e}n waves may undergo a phase
mixing due to the $x$ dependence of the Alfv{\'e}n speed (Tsiklauri
\& Nakariakov 2002). However the aim of this paper is to show the
energy conversion from 3-min oscillations into Alfv{\'e}n waves
only, therefore we do not consider the effect of phase mixing here.

\begin{figure}
\begin{center}
   \includegraphics[width=10cm]{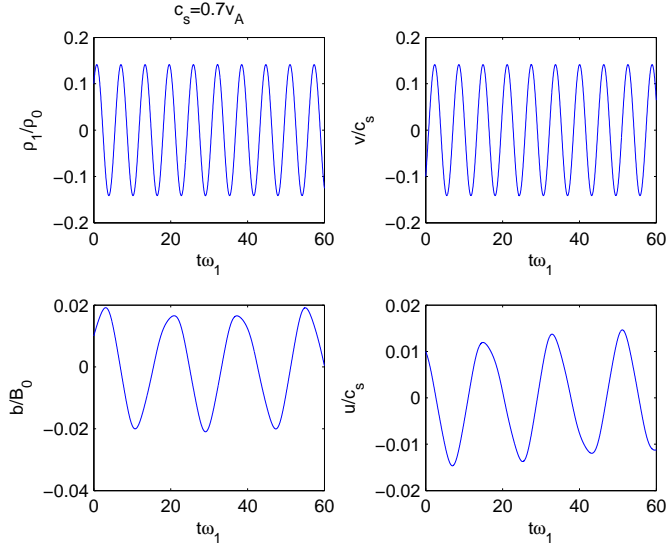}
      \caption{The same as on Fig.1, but for $c_s=0.7v_A$ region. Amplitude of Alfv{\'e}n waves (lower panels)
      shows no increase in time, so there is no wave coupling in this region.}
        \end{center}
         \label{FigVibStab}
   \end{figure}

Numerical solutions of Eqs. (8)-(11) (here the back reaction of
Alfv{\'e}n waves, i. e. the second term of right hand side in Eq.
(10), is again neglected) are presented on Fig.1-2. Figure 1 shows
the wave dynamics in $c_s=v_A$ region and we see the rapid growth of
Alfv{\'e}n waves with twice the period of acoustic oscillations.
However Fig.2 shows the wave dynamics away from the resonant layer
(in $c_s = 0.7v_A$ region) and we see no energy exchange between the
waves. Thus there is a good agreement between numerical results and
analytical solutions.

It must be mentioned that the equilibrium used in this paper is
simplified as gravitational stratification, which is important in
the solar chromosphere, is ignored. The stratification leads to
Klein-Gordon equation for propagating waves with a cut-off for wave
frequencies (Roberts, 2004). The 3-min oscillations and Alfv{\'e}n
waves are above the cut-off, so they may propagate in the
chromosphere. Unfortunately, the stratification greatly complicates
mathematical description of the non-linear coupling. On the other
hand, plasma $\beta$ can be constant along a vertical magnetic tube
even in the case of stratification (Roberts, 2004); this is the case
when sound and Alfv{\'e}n speeds vary with height but their ratio
remains constant. Therefore if the waves are coupled at $\beta \sim
1$ region in an unstratified atmosphere, then the same can be
expected in a stratified medium. However the wave coupling in the
stratified atmosphere needs further detailed study, but is not the
scope of present paper. Our goal is just to show that the 3-min
oscillations may transfer energy into incompressible Alfv{\'e}n
waves in $\beta \sim 1$ region, which was recently observed by
Muglach et al. (2005).

\section{Conclusions}

Here we show that acoustic oscillations trapped in transversally
inhomogeneous medium can be resonantly absorbed by Alfv{\'e}n waves
near the layer of $v_A \approx c_s$. The spatial width of the layer
depends on amplitude of acoustic oscillations and can be
significantly wider for strong amplitude oscillations.

We consider the observed 3-min oscillations as the standing
fundamental harmonic of acoustic waves trapped in the solar
chromospheric cavity with vertical magnetic field and show that
their nonlinear coupling to Alfv{\'e}n waves may take place near
$\beta \sim 1$ layer. The coupling may explain the recent
observational evidence of compressible wave absorption near
$\beta\sim 1$ region of solar lower atmosphere (Muglach et al.,
2005). The amplified Alfv{\'e}n waves with the period of $\sim$ 6
min may carry energy into the corona. There they may deposit the
energy back to density perturbations leading to observed intensity
variations in coronal spectral lines.

Thus the layer of $\beta \sim 1$ may play a role of {\it energy
channel} for otherwise trapped acoustic oscillations guiding the
photospheric energy into the solar corona. Therefore the process of
wave coupling can be of importance for coronal heating problem, but
requires further study especially for a stratified atmosphere.

\section{Acknowledgements}

The work was supported by the grant of Georgian National Science
Foundation GNSF/ST06/4-098 and the NATO Reintegration Grant FEL.RIG
980755. A part of the work is supported by the ISSI International
Programme "Waves in the Solar Corona".

\end{document}